\documentclass[12pt,preprint,flushrt]{aastex}
\usepackage{epsf}

\shortauthors{KOYAMA \& INUTSUKA}

\begin{document}            

\title{Nonlinear Development of Thermal Instability  
without External Forcing}

\author{Hiroshi Koyama\altaffilmark{1} and
Shu-ichiro Inutsuka\altaffilmark{2}}

\altaffiltext{1}{Department of Earth and Planetary System Science, 
Graduate School of Science and Technology,
Kobe University, Nada, Kobe 657-8501, Japan; 
hkoyama@kobe-u.ac.jp}
\altaffiltext{2}{
Department of Physics, Kyoto University, Kyoto 606-8502, Japan;
inutsuka@tap.scphys.kyoto-u.ac.jp}

\begin{abstract}
Supersonic turbulent motions are the remarkable properties of 
 interstellar medium. 
Previous numerical simulations have demonstrated that 
 the thermal instability in a shock-compressed layer produces 
 the supersonic turbulent motion that does not decay. 
In this paper 
 we focus on two- and three-dimensional numerical simulations of 
 the non-linear development of simple thermal instability 
 incorporating physical viscosity but without any external forcing,  
 in order to isolate the effects of various processes responsible 
 for the long-lasting turbulent motion. 
As the initial condition for our simulations, 
 we set up spatially uniform gas with thermally unstable temperature 
 in a box with periodic boundaries. 
After the linear growth stage of the thermal instability, 
 two-phase medium forms where cold clumps are embedded in 
 warm medium, and turbulent fluid flow clearly visible as 
 translational motions of the cold clumps does not decay 
 in a viscous dissipation timescale.
The amplitude of the turbulent velocity increases when we 
 reduce the Prandtl number 
 that is the non-dimensional ratio of kinetic viscosity to 
 thermal conduction:  
the saturation amplitude does not change when we increase  
 the viscosity and thermal conduction coefficients simultaneously 
 in order to keep the Prandtl number.  
This shows that the thermal conduction plays an important role 
 in maintaining turbulent motions against viscous dissipation. 
The amplitude also increases when we increase the ratio of 
 the computational domain length $L$ to 
 the cooling length $\lambda_{\rm c}$ 
 that is defined by the product of the cooling time and 
 the sound speed, 
 as long as $L \la 100 \lambda_{\rm c}$.  
\end{abstract}
\keywords{hydrodynamics --- ISM: general
--- method: numerical --- instabilities, turbulence}

\section{Introduction}
The interstellar medium (ISM) is observed to be highly turbulent
 (Heiles \& Troland 2003). 
Turbulence may play an important
 role in the subsequent evolution of ISM such as the formation of dense
 molecular clouds (e.g., Mac Low \& Klessen 2004).  
A number of hydrodynamical processes 
 have been
 proposed as a mechanism for generating turbulent motions in
 multi-phase ISM:
 supernova explosions (e.g., de Avillez \& Mac Low 2002), 
 Galactic differential rotation (e.g., Wada \& Norman 2001), 
 MRI (e.g., Piontek \& Ostriker 2004), and
 winds from young stars in active star forming clouds 
 (e.g., Li \& Nakamura 2006).
Among them, 
 energetics argument favors
 supernova explosions as the most efficient ultimate source of
 turbulence (e.g., Mac Low \& Klessen 2004).

When the ISM is swept-up by a compressional wave, 
 the gas temperature in the shock-compressed gas once increases by
 shock-heating but 
 eventually decreases because of higher cooling rate in compressed gas.
If the temperature goes down from 6,000 K to 300 K, the gas becomes 
 thermally unstable 
 (Field, Goldsmith \& Habing 1969; Wolfire et al. 1995, 2003).   
The nonlinear evolution of thermal instability (TI) in this situation
 has been studied by Koyama \& Inutsuka (2002, hereafter KI02;
 Inutsuka \& Koyama 2002; Inutsuka \& Koyama 2004).
They have found that TI generates supersonic turbulence which does not
 decay as long as the shock wave continues its propagation. 
In similar context, formation of cold clouds in warm turbulent
 colliding flows 
 is also analyzed by 
Audit \& Hennebelle (2005), Heitsch et al. (2005) and 
V\'azquez-Semadeni et al. (2006). 

In contrast to KI02,
 many numerical studies on compressible turbulence of hydro- and
 magnetohydrodynamics in one-phase medium (i.e., isothermal gas) have
 shown that the kinetic 
 energy decays within a crossing time or less (Stone, Ostriker \&
 Gammie 1998, Mac Low, Klessen \& Burkert 1998, Cho \& Lazarian 2004).    
In case of MHD simulations of supersonic turbulence,
 about half of energy is dissipated by shocks and the rest of it is 
 dissipated by physical or numerical viscosity.
Decaying turbulence in two-phase medium is also reported by Kritsuk
 \& Norman (2002).  
They have reported a decay rate of TI-induced turbulence that is
 similar to that of one-phase.

In order to clarify the importance of TI in the turbulent ISM,
 it is imperative to understand why TI-induced turbulence in KI02 does not
 show rapid decay.
One obvious reason is due to continuous supply of unstable gas
 into the post-shock region from the upwind flow.
In other words, TI in the shock propagating model is
 continously happening in ``fresh'' thermally unstable gas 
 that is continuously provided by  
 shock compression and heating. 
Another aspect of the shock propagation model
 is that a supersonic motion of CNM does not mean 
 supersonic with respect to WNM because 
 two phases have very different sound speeds.   
A cold gas clump that
 has a subsonic velocity with respect to WNM can survive shock
 dissipation in surrounding WNM. 
Piontek \& Ostriker (2004) have found 
 long-lasting subsonic turbulence driven by TI, 
 but have not treated the effect of physical viscosity that 
 is important in this subsonic turbulence as shown below.

In this paper,
 we study the dissipation mechanism in TI-induced turbulence
 by performing two- and three-dimensional hydrodynamical simulations with
 realistic cooling/heating rates, thermal conduction, and physical
 viscosity. 
For the sake of simplicity, magnetic field and self-gravity are not
 included in this study.   
Numerical models and methods are described in \S 2.
In \S 3, we demonstrate the kinetic properties of two-phase medium.
In \S 4, we discuss the implications of our results.
Finally, we summarize our findings in \S 5.

\section{Numerical Methods}

We solve the following compressible Navier-Stokes equations with
 cooling and heating terms: 
\begin{eqnarray}
\frac{\partial\rho}{\partial t} &+& 
\frac{\partial}{\partial x_j}(\rho u_j)=0,\\
\frac{\partial\rho u_i}{\partial t} &+& 
\frac{\partial}{\partial x_j}(\rho u_iu_j-\sigma_{ij})
=-\frac{\partial P}{\partial x_i},\\
\frac{\partial E}{\partial t} &+& 
\frac{\partial}{\partial x_j} 
\left[
(E+P)u_j-\sigma_{ij}u_i-K\frac{\partial T}{\partial x_j} 
\right] \nonumber \\
&&=n\Gamma-n^2\Lambda(T),\\
E &=& \frac{P}{\gamma-1}+\frac{\rho u_ku_k}{2},\\
P &=& n k_{\rm B}T, \\
n &=& \rho/m_{\rm H}, \\
\sigma_{ij} &=& \mu\left\{
\left(
\frac{\partial u_i}{\partial x_j}+\frac{\partial u_j}{\partial x_i}
\right)
-\frac{2}{3}\delta_{ij}\frac{\partial u_k}{\partial x_k}\right\}.
\end{eqnarray}
All symbols have their usual meanings. 

We take the realistic cooling function $\Lambda(T)$,
\begin{eqnarray}
 \frac{\Lambda(T)}{\Gamma}=1.0\times 10^7 \exp(-1.184\times 10^5/(T+1000))
+14\times 10^{-3} \sqrt{T} \exp(-92/T)
\end{eqnarray}
 and the density independent heating rate $\Gamma$
 from KI02\footnote{There were two typos in eqn (4) of KI02.}.
We assume the classical conduction coefficient of neutral hydrogen,
 $K = 2.5 \times 10^3 T^{1/2}\,{\rm cm^{-1} K^{-1} s^{-1}}$
 (Parker 1953) for a fiducial model. 
The relation between the viscosity $\mu$ and the thermal conductivity $K$
 is characterized by Prandtl number:
\begin{equation}
{\rm Pr}=
\frac{\gamma}{\gamma-1}\frac{k_{\rm B}}{m_{\rm H}}\frac{\mu}{K},
\end{equation}
where $\gamma$ is the ratio of specific heats, $k_{\rm B}$ is the
 Boltzmann constant and $m_{\rm H}$ is the mass of atomic hydrogen.
Note that the Prandtl number for a neutral monoatomic gas is Pr=2/3.
Throughout this paper 
 we assume a gas consists of atomic hydrogen ($\gamma=5/3$).

We use an operator-splitting technique for solving the equations:
 the second-order Godunov method (van Leer 1979) for inviscid fluid part
 and first-order explicit time integration for cooling, heating, thermal
 conduction and physical viscosity parts.
The conduction and viscosity operators have spatially second-order
 accuracy. 
The time step in our code is determined by the three criteria:
(1) CFL condition, 
$\Delta t_{\rm CFL}\le C_{\rm CFL}\Delta x/(c_s+|v_i|)$ 
with $C_{\rm CFL}=0.5$, 
(2) cooling and heating timescales, 
$\Delta t_{\rm cool} \le 0.2\frac{k_{\rm B}T}{\gamma-1}{\rm
 min}(\frac{1}{\Gamma},\frac{1}{n\Lambda(T)})$, and 
(3) conduction and viscosity timescales, 
$\Delta t_{\rm vis} \le 0.2{\rm min}
(\frac{nk_{\rm B}}{(\gamma-1)K},\frac{\rho}{\mu} )(\Delta x)^2
=0.2\frac{nk_{\rm B}}{(\gamma-1)K}{\rm min}(1,\gamma/{\rm Pr})(\Delta x)^2$.
We use uniform Cartesian grids with periodic boundary condition.

\section{Characteristic Scales of TI}

\begin{table}
\begin{center}
\caption{Initial parameters and characteristic lengths for the fiducial
 model}
\begin{tabular}{rr}\tableline\tableline
Parameter & Value \\
\tableline 
$n_0$\tablenotemark{a} (cm$^{-3}$) & 4.3 \\
$T_0$ (K) & 423 \\
$P_0/k_{\rm B}$ (K cm$^{-3}$) & 1820 \\
$\lambda_{\rm F}$\tablenotemark{b} (pc) & 0.0054 \\
$\lambda_{m}$\tablenotemark{c} (pc) & 0.044 \\
$\lambda_{c}$\tablenotemark{d} (pc) & 0.35 \\
\tableline
\end{tabular}
\tablenotetext{a}{Initial number density.
Note that this is equal to the mean number density.}
\tablenotetext{b}{Field length}
\tablenotetext{c}{The most unstable wavelength}
\tablenotetext{d}{Cooling length}
\end{center}
\end{table}

The development of TI is characterized by three length scales:   
 the critical length scale, the
 cooling length, and the most unstable wavelength.   
We briefly summerize those length-scales in this section. 

Even when the gas is thermally unstable,
 the thermal conduction can erase the perturbation
 whose wavelength is sufficiently small.
In other words, there is a critical wavelength of TI called
 `Field length' which is defined by
\begin{eqnarray}
\lambda_{\rm F} &=& \sqrt{\frac{KT}{n^2\Lambda}} \nonumber \\
&=& 1.4\times 10^{-2} \frac{T_2^{0.35}}{n} ~ {\rm pc}
\end{eqnarray}
 (Field 1965) where $T_2=T/(10^2{\rm K})$.
In this expression, we adopt a simple power law cooling rate 
 (eqn. [3] in Wolfire et al. 2003).
Note that this cooling rate shows a good agreement with the detailed 
 calculations 
 in the temperature range from 100 to 8,000 K 
(e.g., Wolfire et al. 1995, 2003 and Koyama \& Inutsuka 2000). 
The thermal conduction is important in the two-phase interface region
 at which radiative cooling/heating balances conductive
 heating/cooling (Inutsuka, Koyama \& Inoue 2005).  
Graham and Langer (1973) studied cold spherical clouds embedded in
 warm gas and found that small clouds can evaporate due to this
 conduction while ambient warm gas can condenses on large clouds.
Obviously, the transition region of the two-phase interface must be
 resolved to correctly describe those evaporation and condensation of clouds.
The thickness of this region is the order of the Field
 length (Begelman \& McKee 1982). 
Therefore, the Field length should be resolved in numerical analysis
 of TI.
We termed this criterion as the `{\it Field condition}' 
 (Koyama and Inutsuka 2004, hereafter KI04).

Next, we introduce the cooling length given by
\begin{eqnarray}
\lambda_c=c_s \tau_{\rm cool}
&=&
\sqrt{\frac{\gamma k_{\rm B}T}{m_{\rm H}}}
\frac{1}{\gamma-1}\frac{k_{\rm B}T}{n\Lambda} \nonumber \\
&=& 0.55 \frac{T_2^{0.7}}{n} ~ {\rm pc},
\end{eqnarray}
where $\tau_{\rm cool}$ is the cooling time and $c_s$ is the sound speed.
In the scale smaller than this, 
 pressure balance sets in rapidly on the 
 sound crossing time which is less than the cooling time.
In other words, the perturbation evolves almost isobarically.
Alternatively, in the long-wavelength limit, the sound crossing time
 becomes relatively long, and radiative equilibrium sets in first.
The radiative equilibrium pressure depends on the local density
 (see, e.g., solid line in {\it right} panel of Figure \ref{L4}).
This means that the density gradient produced by TI also 
 produces pressure gradient
 that can produce larger dynamical motions than isobaric motions.
S\'anchez-salcedo et al. (2002)
 showed that large-scale perturbations generated 
 large condensation flows with Mach numbers larger than
 0.5, but small-scale perturbations evolved less dynamically.

Finally, we introduce the most unstable wavelength of TI 
 which is approximately expressed by the geometric mean of the two
 characteristic lengths
\begin{equation}
\lambda_m \approx\sqrt{\lambda_{\rm F}\lambda_c}
=0.088\frac{T_2^{0.525}}{n} ~ {\rm pc}
\end{equation}
(Field 1965).
Note that we have a size relation 
\begin{eqnarray}
\lambda_{\rm F} < \lambda_{m} < \lambda_c,
\end{eqnarray}
where the realistic parameters are assumed.
These three characteristic lengths of the fiducial model are listed in Table 1.

In order to simulate a gas flow of two-phase medium such as 
 condensation and evaporation of clouds and evolution of a thermally
 unstable mode,  
 at least the Field length $\lambda_{\rm F}$ and the most unstable
 wavelength $\lambda_m$ should be resolved.
In addition, the cooling length may be important for the dynamical
 evolution of TI.  
Since $\lambda_c$ is about two order of magnitude larger than
 $\lambda_{\rm F}$,
 at least a few hundreds of grids per one-dimension are required to
 resolve those length-scales in a numerical analysis.

\section{Nonlinear Evolution of TI}

In this section, we investigate 
 the development of TI, in particular, production of
 supersonic turbulence and energetics of two-phase turbulence. 

\subsection{TI-Induced Turbulence}

\begin{figure}
\figurenum{1}
\epsscale{1.0}
\plotone{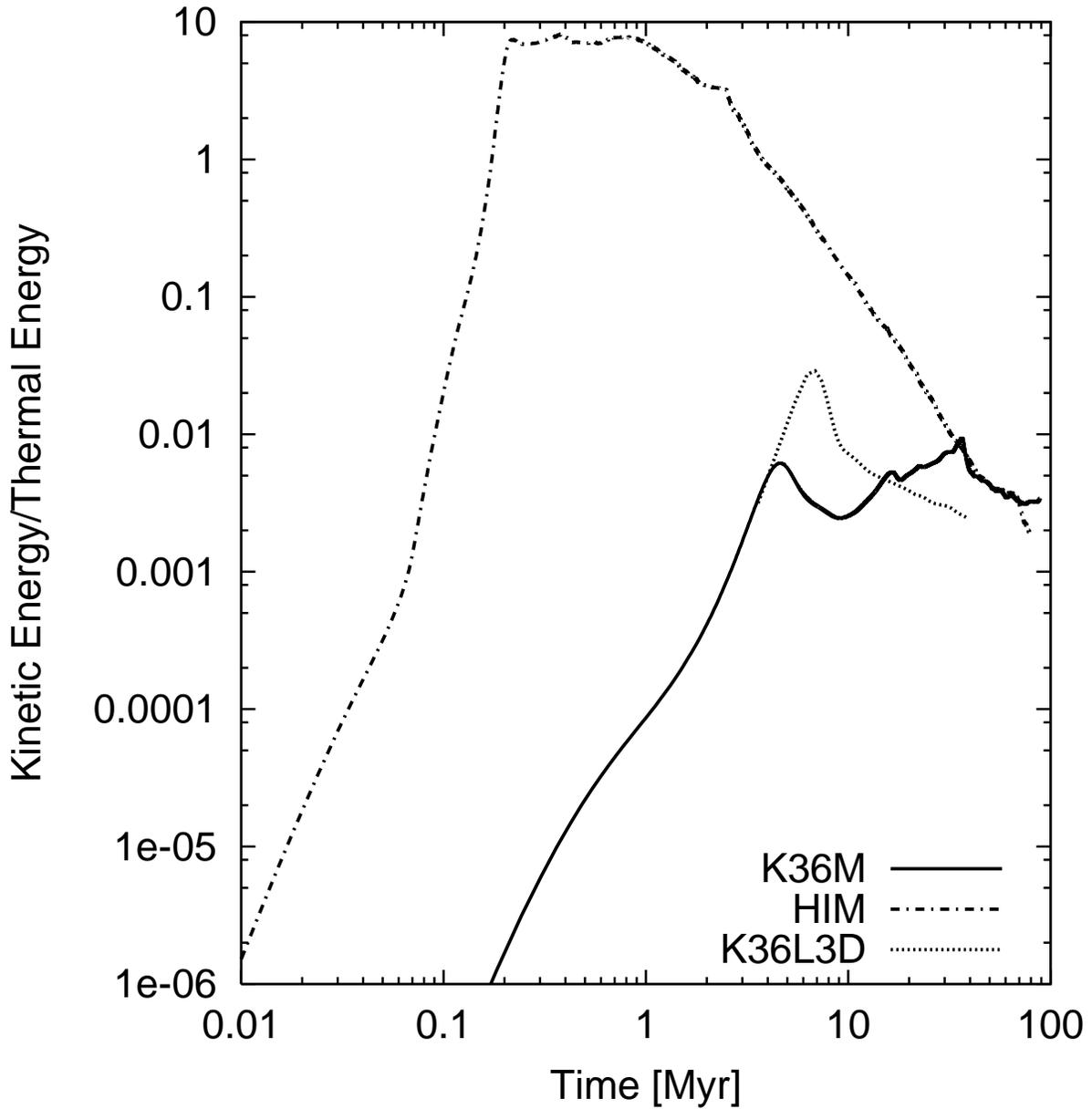}
 \caption{The ratio of kinetic to thermal energies.
{\it Dot-dashed line}: two-dimensional model with initially hot medium
 (5$\times 10^5$ K). 
{\it Solid line}: two-dimension with initially unstable one-phase
 medium (Model K36M, see Table 2). 
{\it Dotted line}: same as the Model K36M but three-dimension
 (Model K36L3D).}
\label{HIM}
\end{figure}

As for initial conditions,
 we assume uniform gas with small isobaric density perturbations 
 so that the development of velocity fluctuations is purely caused by TI. 
Figure \ref{HIM} shows time evolution of the ratio of kinetic to
 thermal energies for the three different models.
Dot-dashed line denotes the two-dimensional simulation starting with hot
 medium ($5\times 10^5$ K). 
The kinetic energy produced by TI exceeds thermal energy at 
 $\sim$0.1 Myr which corresponds to a few cooling time of hot gas.
This temporal turbulence eventually decays.
This decay property in multiphase ISM was reported by
 Kritsuk \& Norman (2002). 

\begin{figure}
\figurenum{2}
\epsscale{1.0}
\plotone{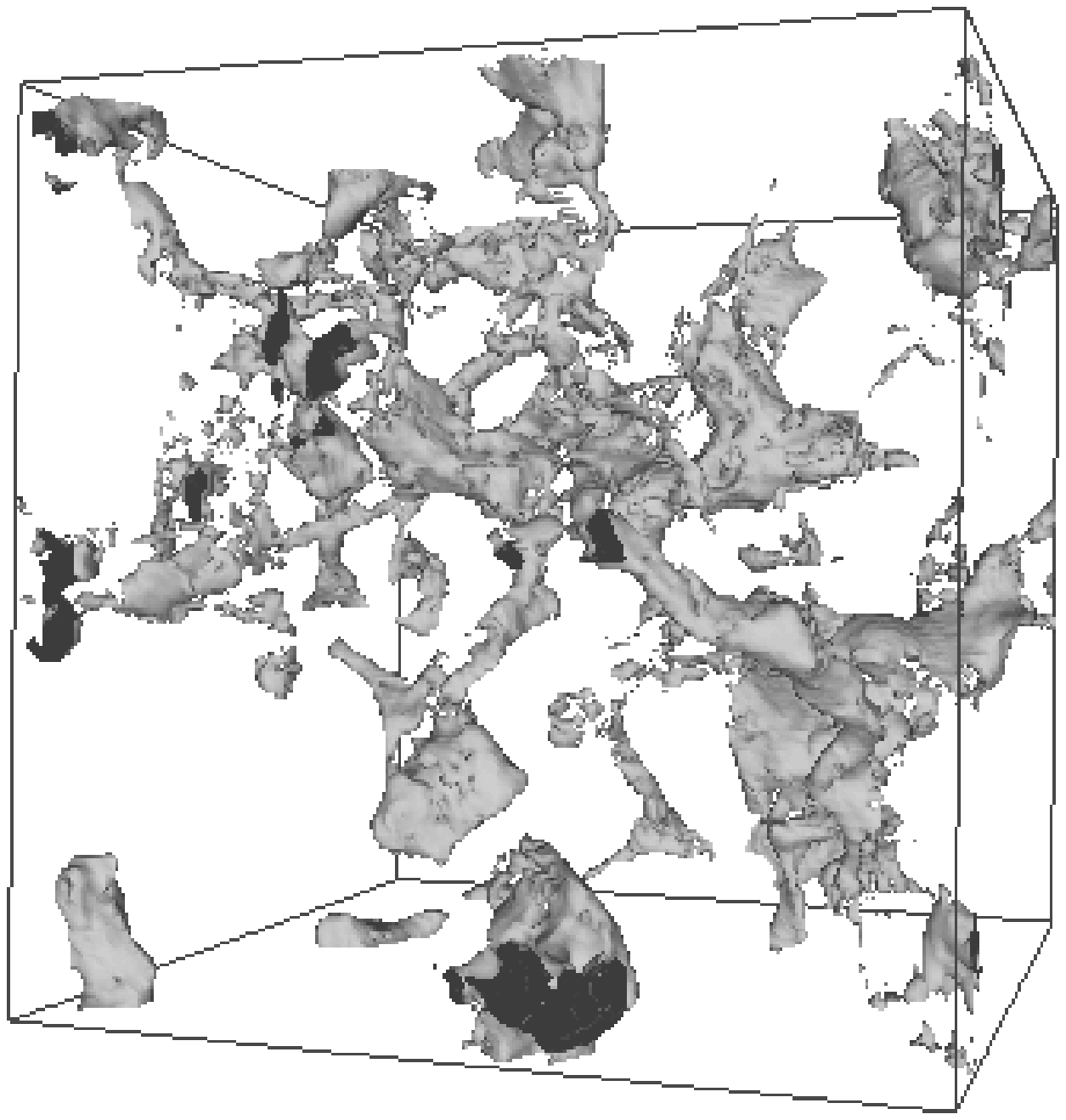}
 \caption{Iso-density surface for Model K36L3D (L=36 pc). 
128$^3$ grids are used.}
\label{3D}
\end{figure}

Solid and dotted lines in Figure \ref{HIM} show the simulations
 starting with unstable one-phase 
 medium in two- and three-dimension, respectively.
This unstable initial condition is widely used for the simulation of
 TI (e.g., S\'anchez-salcedo et al. 2002; Piontek \& Ostriker 2004; KI04).
We could not find any significant difference between two- and
 three-dimensional models.
One snapshot (three-dimensional model K36L3D) is
 presented in Figure \ref{3D}.
The increasing kinetic energy attains maximum at $\sim$ 5 Myr 
 which corresponds to a cooling time of WNM.
After the linear growth stage of TI, non-linear turbulent state is
 established, irrespective of the difference in the initial condition. 
We call this dynamical equilibrium state ``saturation.''      
In this saturated stage, the kinetic energy is about 1\% of the
 thermal energy.
It is interesting to note that these energies in the three different initial
 models converge on the same level.
This seems that the saturation of turbulence driven by TI is one of
 the basic hydrodynamical property of the two-phase medium 
 independent on the initial condition.

In order to satisfy the Field condition, 
We use artificially large conduction coefficients 
 in these three models. 
The corresponding viscosity coefficients are determined by fixing the
 Prandtl number 2/3 except for Model K36L3D.  
The dependence of the saturation on the conduction and viscosity
 coefficients is described in the subsequent sections.

\subsection{Fiducial Model of the TI-induced Turbulence}

\begin{figure}
\figurenum{3}
\epsscale{0.7}
\plotone{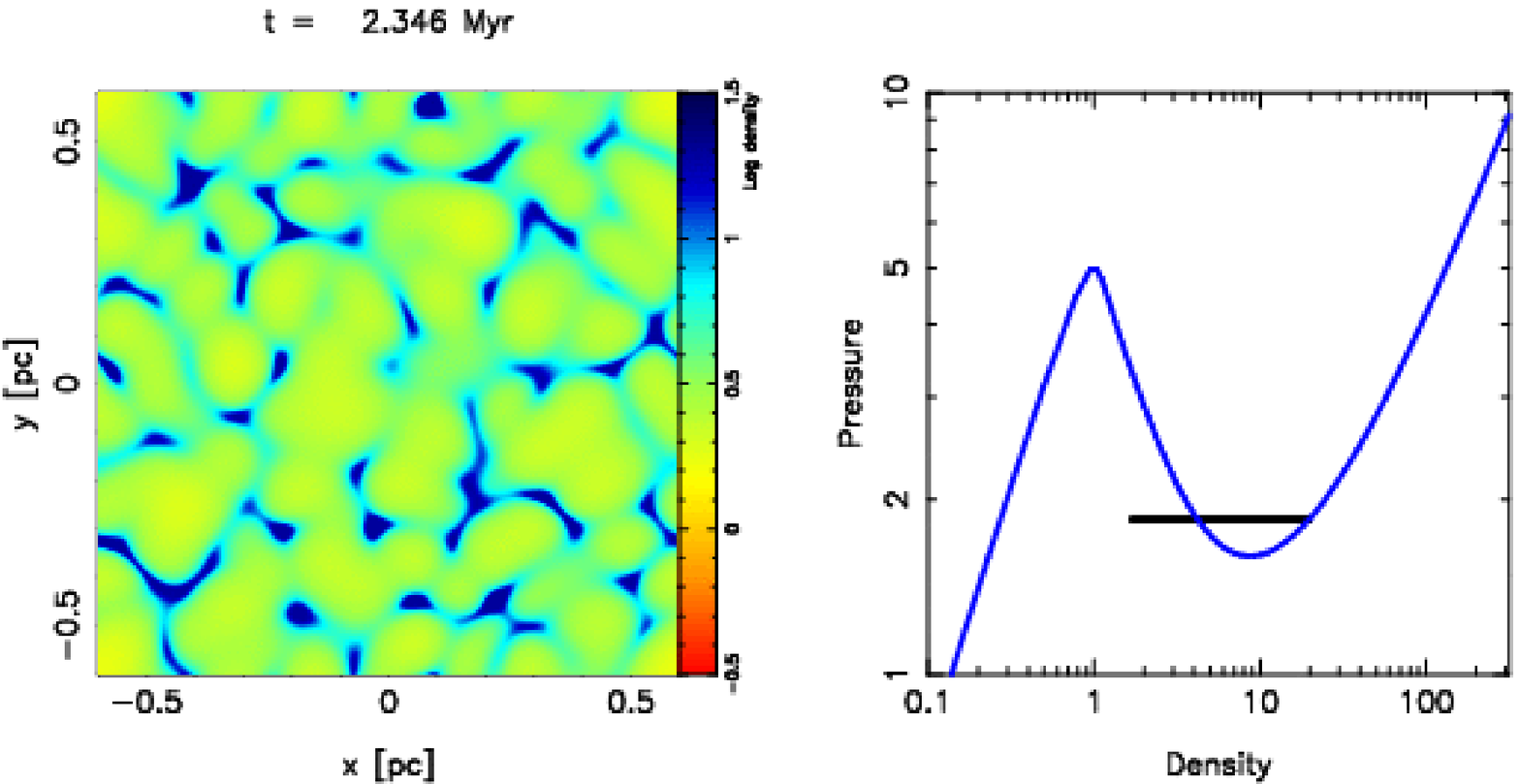}
\plotone{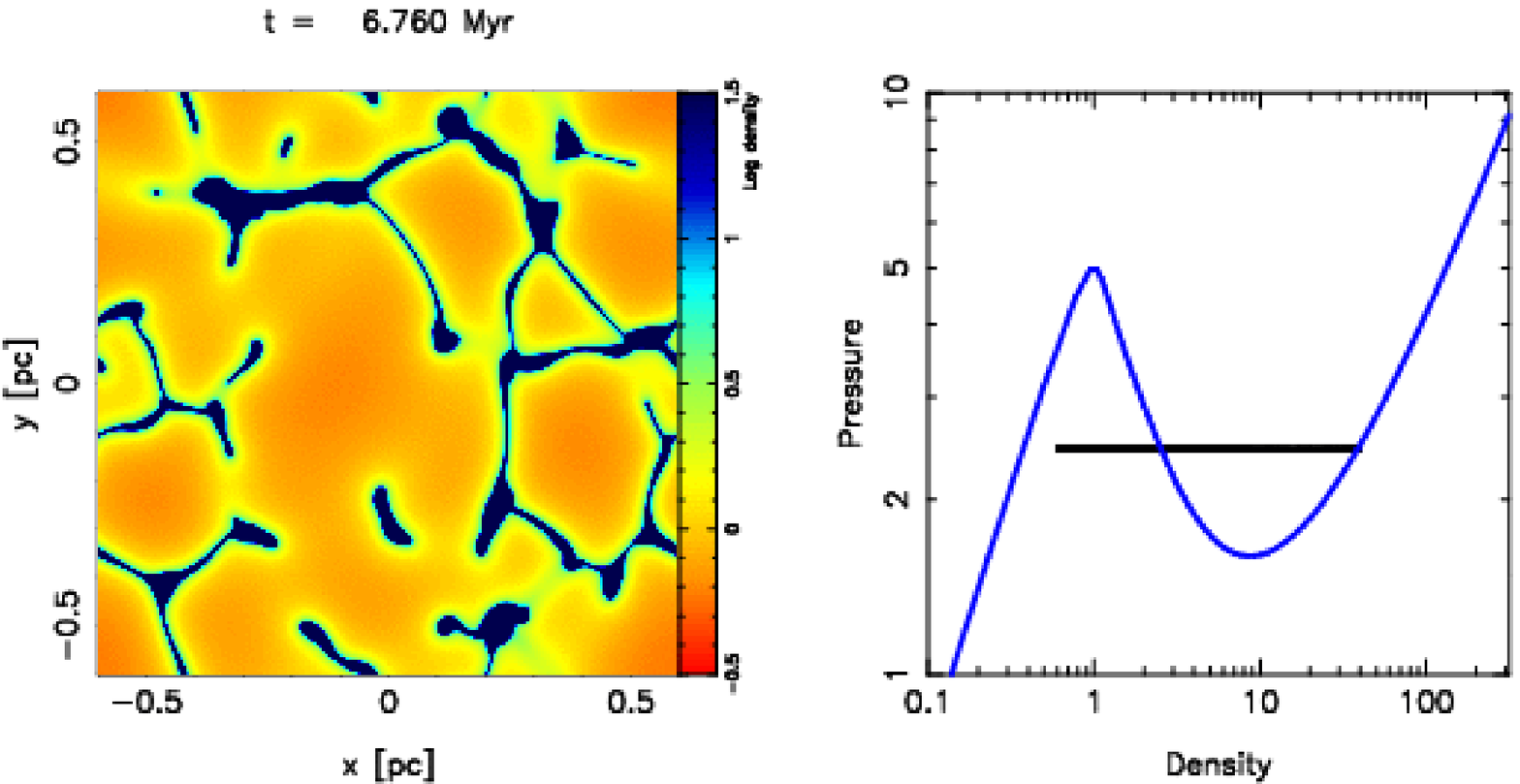}
\plotone{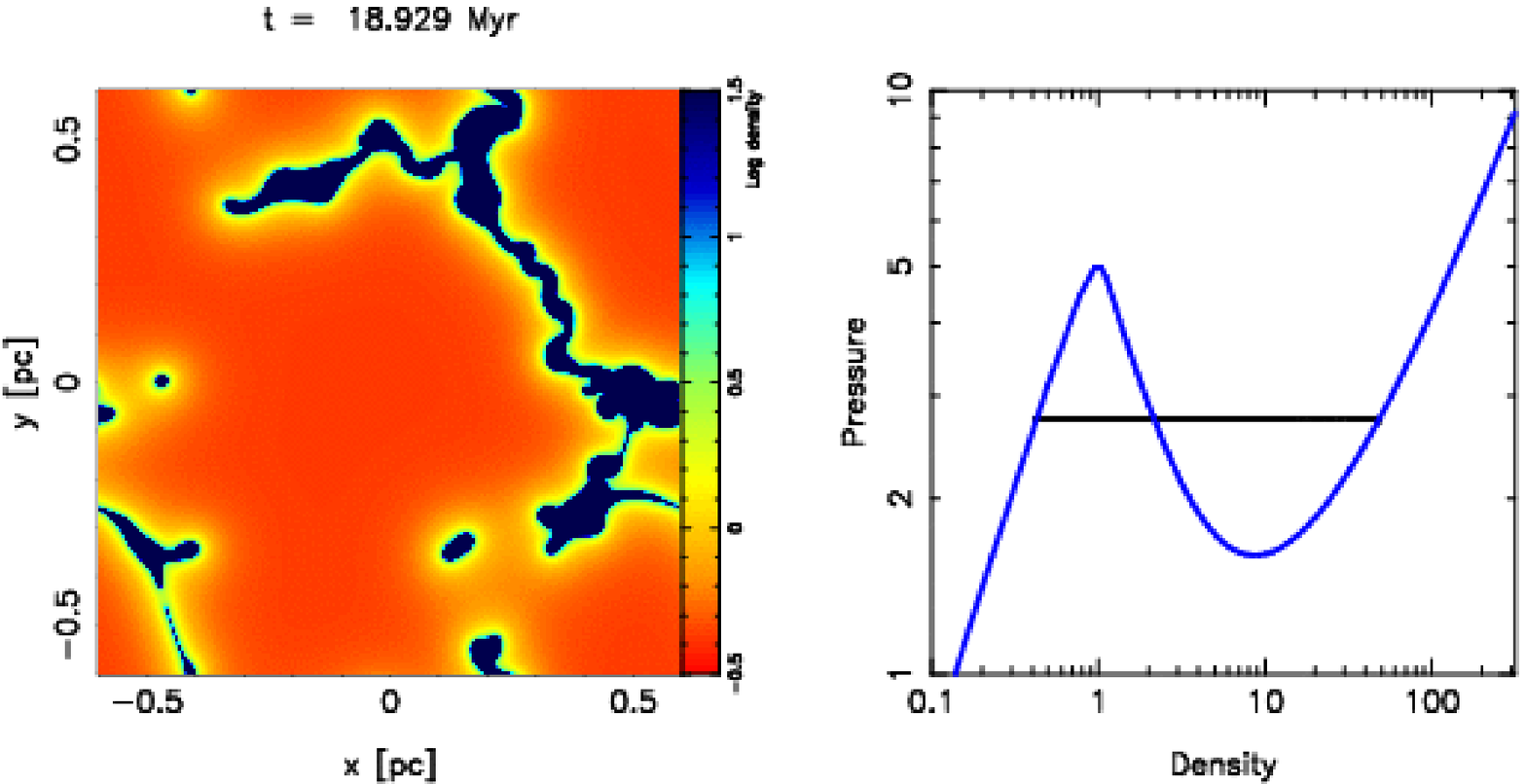}
 \caption{
Evolution of TI for Model L4S (Pr=2/3, L=1.2 pc): 
{\it Top}: Early stage.
Over density perturbations grow and reach the CNM
 phase while less dense perturbations have been growing toward the
 radiative equilibrium WNM phase. 
{\it Middle}: 
Intermediate stage. 
Less dense medium almost attains the radiative equilibrium WNM phase.
The spatially uniform pressure evolves gradually.
{\it Bottom}: 
Saturated stage. 
Small clouds evaporate and large clouds condense continuously.}
\label{L4}
\end{figure}

Overall evolution of TI-induced
 turbulence in two-dimension with realistic dissipation coefficients
 is presented in this subsection.
Initial parameters of the unperturbed state are listed in Table 1.
The domain length (1.2pc) is 3.4
 times larger than the cooling length $\lambda_c$ so that
 long-wavelength mode of TI can develop.  

Figure \ref{L4} shows three snapshots of the density
 distribution alongside scatter plots of the pressure-density distribution.
The radiative equilibrium curve as a function of given density is also
 plotted as a blue line. 
Top panels exhibit the early stage of the TI evolution.
Initially over density regions grow and attain the equilibrium state
 of CNM while less density regions have not attained  
 the WNM phase at that moment because of a longer heating timescale.
Since the most unstable wavelength $\lambda_m$ is less than the
 cooling length $\lambda_c$,
 the fastest growing mode is almost isobaric:   
 the spatially uniform pressure is nearly the same as in the
 initial value (see also Figure \ref{ene}a).

Middle panels in Figure \ref{L4} show the stage that the dilute
 gas approaches the WNM phase.
Toward the WNM phase, the gas must expand in order to reduce its density.
Owing to the fixed volume boundaries, expanding warm gas compresses the
 cold dense clouds as well as warm gas itself. 
Consequently, the gas pressure slightly increases by forming WNM phase.
The time evolution of the gas pressure is shown in Figure
 (\ref{ene}a).  

Bottom panels in Figure \ref{L4} show fully developed stage of
 two-phase medium.  
Both evaporation of the small clouds and accretion of the surrounding
 warm gas onto the cold clouds are seen.

Since the clouds of the size smaller than $\lambda_{\rm F}$ must be
 erased by thermal conduction,
 the minimum size of cold clouds is the order of $\lambda_{\rm F}$.
For those small clouds, 
 thermal conduction provides a significant heating so that
 the clouds evaporate by the conductive heating.
Neglecting the radiative heating and cooling terms, integration of the
 energy equation gives the following equation:
\begin{equation}
\frac{d}{dt}\int_{V} \frac{P}{\gamma-1} dV 
=\int_{S} K\nabla T\cdot d{\bf r}
\sim S\frac{KT}{\lambda_{\rm F}},
\end{equation}
where $V$ and $S$ are volume and surface of an isolated cloud, respectively.
In the last part, we use the approximation that the temperature
 gradient is the order of $T/\lambda_{\rm F}$.
This equation leads to the evaporation timescale
\begin{equation}
\tau_{\rm evap}\sim
\frac{V}{S}\frac{P\lambda_{\rm F}}{(\gamma-1)KT}
= \tau_{\rm cool}\frac{R}{2\lambda_{\rm F}}.
\end{equation}
Here, we adopt cylindrical symmetry for
 the two-dimensional simulations, using the relation $V/S=R/2$ where $R$
 is the radius of the cloud.
This equation indicates that a small cloud can evaporate in a few cooling
 time. 
Detailed analysis of cloud evaporation are presented
 in Nagashima, Koyama \& Inutsuka (2005) and Nagashima, Inutsuka \&
 Koyama (2006).

\subsection{Energy Budget and Dissipation Timescale}

\begin{figure}
\figurenum{4}
\epsscale{0.7}
\plotone{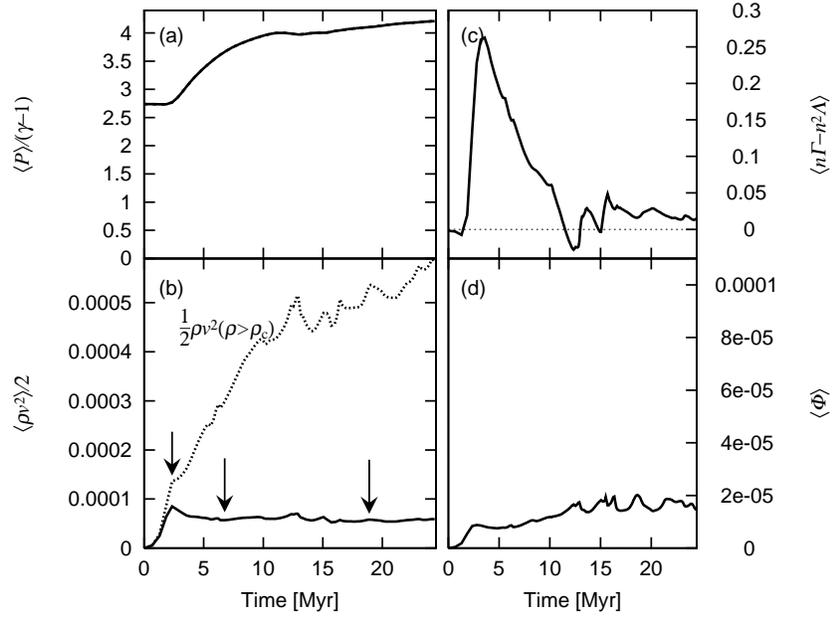}
 \caption{Evolution of (a) thermal  
 and (b) kinetic energies in units of 
 10\protect{$^{-3} k_{\rm B}$} ergs cm\protect{$^{-3}$}.  
%1.38 x 10\protect{$^{-13}$} ergs cm\protect{$^{-3}$}.  
 Models L4S is plotted (Pr=2/3, L=1.2pc. See Table 2). 
Three arrows correspond to the snapshots in Figure \ref{L4}. 
Dotted line (panel b) shows the kinetic energy of the high density gas
 ($\rho>\rho_c=4$ cm\protect$^{-3}$).
Panel c and d denote the net radiative and viscous heating in unit of
 \protect{$2\times 10^{-26}$ ergs s$^{-1}$}, respectively. 
All quantities are the volume averages.  }
\label{ene}
\end{figure}

Figure \ref{ene}a and b exhibit the time evolution of volume averaged
 thermal and kinetic energy, respectively. 
The three arrows plotted in Figure \ref{ene}b correspond to the
 snapshots in Figure \ref{L4}.
The thermal energy dominates in this model.
The kinetic energy does not decay but shows saturation.
The kinetic energy of high density gas (Dotted line in Figure 4b)
 is much larger than the averaged value (Solid line).

Owing to the periodic boundary condition, 
 the total energy $E$ (thermal plus kinetic energies per unit volume)
 varies only with the net heating strength.
The time evolution of the total energy is given by
\begin{eqnarray}
\frac{\partial}{\partial t}\langle E\rangle &=& 
\langle n\Gamma-n^2\Lambda(T)\rangle \nonumber \\
&=& \bar{n}\Gamma - \langle n^2\Lambda(T)\rangle,
\label{eq:ave}
\end{eqnarray}
where the bracket denotes the volume average.  
Since we adopt 
 the constant heating $\Gamma$ and the total mass is conserved, 
 the total thermal energy changes according to the constant heating
and the variable cooling rate $\langle n^2\Lambda(T)\rangle$
 that depends on the local density and temperature distribution. 
The net heating rate, therefore, depends on the structure of
 the two-phase medium.
Figure \ref{ene}c shows the net radiative heating rate 
(rhs in eqn.[\ref{eq:ave}])
 as a function of time.
The net gain rate is only on the order of 0.1 \% of the constant
 (average) heating
 rate $\bar{n}\Gamma=8.6\times 10^{-26}$ ergs s$^{-1}$. 
In other words, 
 departure from the overall radiative equilibrium is small.

Owing to viscous dissipation,
 a fraction of kinetic energy continuously converts into thermal energy.
The conversion rate (in two-dimension) is 
\begin{eqnarray}
\Phi&=&2\mu\left[
 \left(\frac{\partial v_x}{\partial x}\right)^2
+\left(\frac{\partial v_y}{\partial y}\right)^2
\right]
+\mu\left(
\frac{\partial v_x}{\partial y}+\frac{\partial v_y}{\partial x}
\right)^2 \nonumber \\
&& -\frac{2}{3}\mu\left(
\frac{\partial v_x}{\partial x}+\frac{\partial v_y}{\partial y}
\right)^2.
\end{eqnarray}
As seen in Figure 4d, dissipation rate is approximately constant after
a few Myr.  
The dissipation timescale is estimated as
$T_{diss}=\langle \rho v^2 /2\rangle/\langle \Phi\rangle \approx 6$
Myr.
Because the kinetic energy does not decrease in this timescale,
 we should conclude that the kinetic energy must be continously supplied
 by some processes.  
The pressure gradients, caused by the inhomogeneous cooling due to
 local density and temperature variation,
 evidently produce the translational motions.
This is analyzed in \S 6.2.

\subsection{Resolution Dependence}

\begin{figure}
\figurenum{5}
\epsscale{0.7}
\plotone{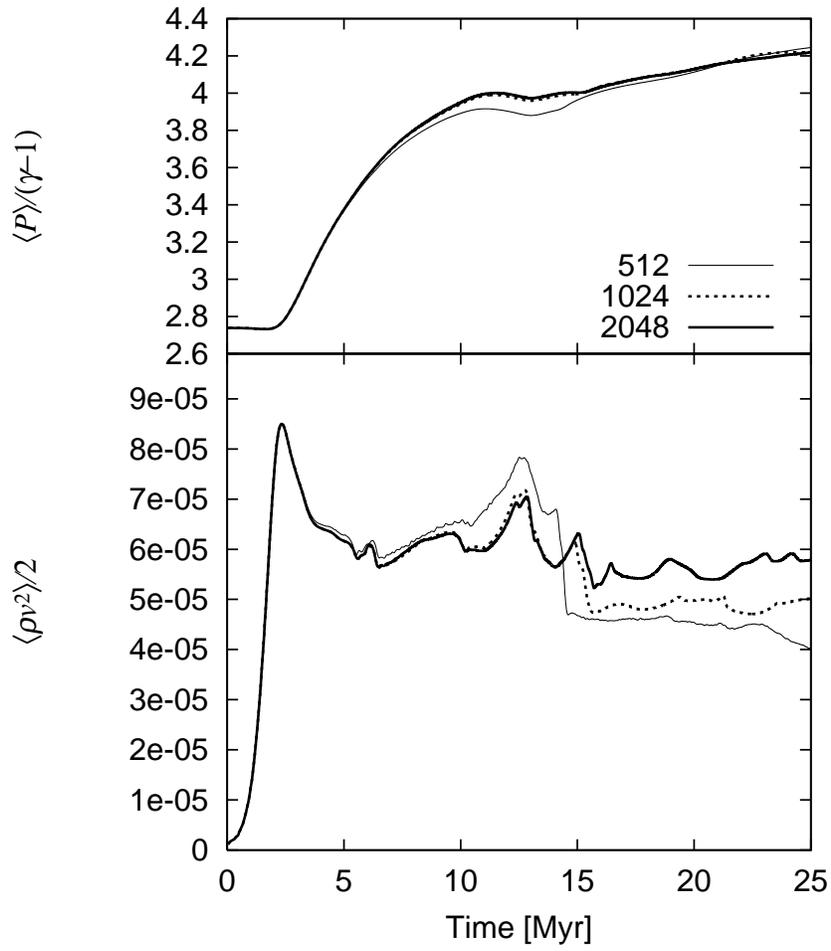}
 \caption{Resolution study. Thermal energy 
$\langle P\rangle/(\gamma-1)$ ({\it Top}) and kinetic energy 
$\langle\rho v^2\rangle/2$ ({\it Bottom}).
Model L4M, L4H and L4S (512, 1024 and 2048 grids in one
 dimension, respectively).}
 \label{resolution}
\end{figure}

The Field length $\lambda_{\rm F}$, the smallest of the three
 characteristic lengths of TI, should be resolved for the simulation
 of TI.
One-dimensional numerical analysis has shown that at least the Field
 number, defined by $N_{\rm F}=\lambda_{\rm F}/\Delta x$, 
 should be greater than three to achieve convergence (KI04).
The fiducial two-dimensional model presented here 
 successfully satisfies this condition by $N_{\rm F}=5.12$ where 2048
 grids in one-dimension are used. 
In Figure \ref{resolution}, we superpose the results of the fiducial model 
 with three different resolutions (i.e., 512$^2$, 1024$^2$ and
 2048$^2$ grids).  
The corresponding Field numbers are 1.28, 2.56 and 5.12,
 respectively. 
When the Field length is moderately resolved 
 (e.g., $N_{\rm F}$=1.28 in Model L4M, see Thin line in Figure
 \ref{resolution}), 
 the kinetic energy is about 20 \% less than that in the high
 resolution Model (L4S) at 25 Myr. 
Because of high numerical cost,
 we use the moderate resolution models
 ($N_{\rm F}\sim 1.28$) to discuss the saturation level.

\section{Saturation of Turbulent Energy}

In the previous section, 
 we demonstrated that the turbulence driven by TI does not decay
 within a viscous dissipation timescale and shows saturation although the
 amplitude of the velocity in the fiducial model (Model L4S) is subsonic. 
It is interesting to know how large can the turbulent velocity grow.
One possibility is, as Sanchez-Salcedo et al. (2000) have been pointed out, 
 that the longer wavelength mode than cooling length
 $\lambda_c$ can develop
 dynamically and consequently produces large turbulent velocity.
The key parameter is thus the ratio of the domain length to cooling
 length, $L/\lambda_c$ which is equivalent to the ratio of sound
 crossing time to cooling time.
For convenience, we refer to the former definition.   
To understand the dependence on the ratio quantitatively, we have
 computed numerous 
 models with different domain lengths $L$ and cooling length
 $\lambda_c$ that are investigated in the subsequent subsections.

\begin{table}
\begin{center}
\caption{Models}
\begin{tabular}{lrrrrrrrr}
\tableline\tableline
Models & Resolutions\tablenotemark{a}
& L(pc) 
& $k_{max}$\tablenotemark{b}  
& $\frac{\lambda_{\rm F,min}}{0.003 ~{\rm pc}}$
& $N_{\rm F}$\tablenotemark{c} 
& Pr 
& $\frac{\lambda_c}{0.35 ~ {\rm pc}}$
&  $\frac{1}{2}\langle\rho v^2\rangle$\tablenotemark{d} \\
\tableline 
L1H          &  256$\times$256 & 0.3 & 4 &   1 & 2.56 & 2/3 &   1 & decay \\
L4M          &  512$\times$512  & 1.2 & 8 &   1 & 1.28 & 2/3 &   1 & 6.0e-5 \\
L4H          & 1024$\times$1024 & 1.2 & 8 &   1 & 2.56 & 2/3 &   1 & 6.0e-5 \\
L4S(fiducial)& 2048$\times$2048 & 1.2 & 8 &   1 & 5.12 & 2/3 &   1 & 6.0e-5 \\
L8M       & 1024$\times$1024 & 2.4 & 8 &   1 & 1.28 & 2/3 &   1 & 2.0e-4 \\
L16M      & 2048$\times$2048 & 4.8 & 8 &   1 & 1.28 & 2/3 &   1 & 5.0e-4 \\
\tableline
K3M   & 512$\times$512  & 3.6 & 8 &   3 & 1.28 & 2/3 &   1 & 5.0e-4 \\
K12M  & 512$\times$512  &  12 & 8 &  10 & 1.28 & 2/3 &   1 & 5.0e-3 \\
K12H  &1024$\times$1024 &  12 & 8 &  10 & 2.56 & 2/3 &   1 & 5.0e-3 \\
K9M   & 128$\times$128           &  9 & 4 &  30 & 1.28 & 2/3 &   1 & 2.0e-3\\ 
K9M3D & 128$\times$128$\times$128&  9 & 4 &  30 & 1.28 & 2/3 &   1 & 2.0e-3\\ 
K36M  & 512$\times$512  &  36 & 8 &  30 & 1.28 & 2/3 &   1 & 2.0e-2 \\ 
K144M &2048$\times$2048 & 144 &16 &  30 & 1.28 & 2/3 &   1 & 2.0e-2 \\ 
\tableline
C5        &  512$\times$512 & 1.2 & 8 &  1 & 1.28 & 2/3 & 0.2   & 1.0e-3  \\
C10       &  512$\times$512 & 1.2 & 8 &  1 & 1.28 & 2/3 & 0.1   & 4.0e-3  \\
C30       &  512$\times$512 & 1.2 & 8 &  1 & 1.28 & 2/3 & 0.033 & 2.0e-2  \\
\tableline
 K1P20     &  512$\times$512 & 1.2& 8 &  1  & 1.28 & 0.05  &  1 & 6.0e-3  \\
 K1P100    & 1024$\times$1024& 1.2& 8 &  1  & 2.56 & 0.006 &  1 & 2.4e-2  \\
 K3P20     &  512$\times$512 & 3.6& 8 &  3  & 1.28 & 0.05  &  1 & 1.5e-2  \\
 K3P100    & 1024$\times$1024& 3.6& 8 &  3  & 2.56 & 0.006 &  1 & 3.3e-2 \\
 K12P20    &  512$\times$512 & 12 & 8 &  10 & 1.28 & 0.05  &  1 & 2.4e-2  \\
 K12P100   & 1024$\times$1024& 12 & 8 &  10 & 2.56 & 0.01  &  1 & 5.0e-2  \\
K36L3D&128$\times$128$\times$128&36&4 &  30 & 0.32 & 0.1   &  1 & 2.0e-2  \\
K144P100   & 2048$\times$2048&144 &16 &  30 & 1.28 & 0.006 &  1 & 6.0e-2  \\
\tableline
HIM\tablenotemark{e}
       & 256$\times$256  & 36 & 4 &  30 & 0.64 & 2/3   &  1 & 2.0e-2  \\
\end{tabular}
\tablenotetext{a}{Number of grid points.}
\tablenotetext{b}{Perturbation wave number.}
\tablenotetext{c}
{The Field number, $N_{\rm F}$ defined by $\lambda_F/\Delta x$}
\tablenotetext{d}{Saturated kinetic energy 
in units of $1.38\times 10^{-13}$ ergs cm$^{-3}$.}
\tablenotetext{e}{
Initially hot medium ($T=5\times10^{5}$ K).}
\end{center}
\end{table}

\subsection{Saturation Levels: Dependence on Box Size}

\begin{figure}
\figurenum{6}
\epsscale{1.0}
 \plotone{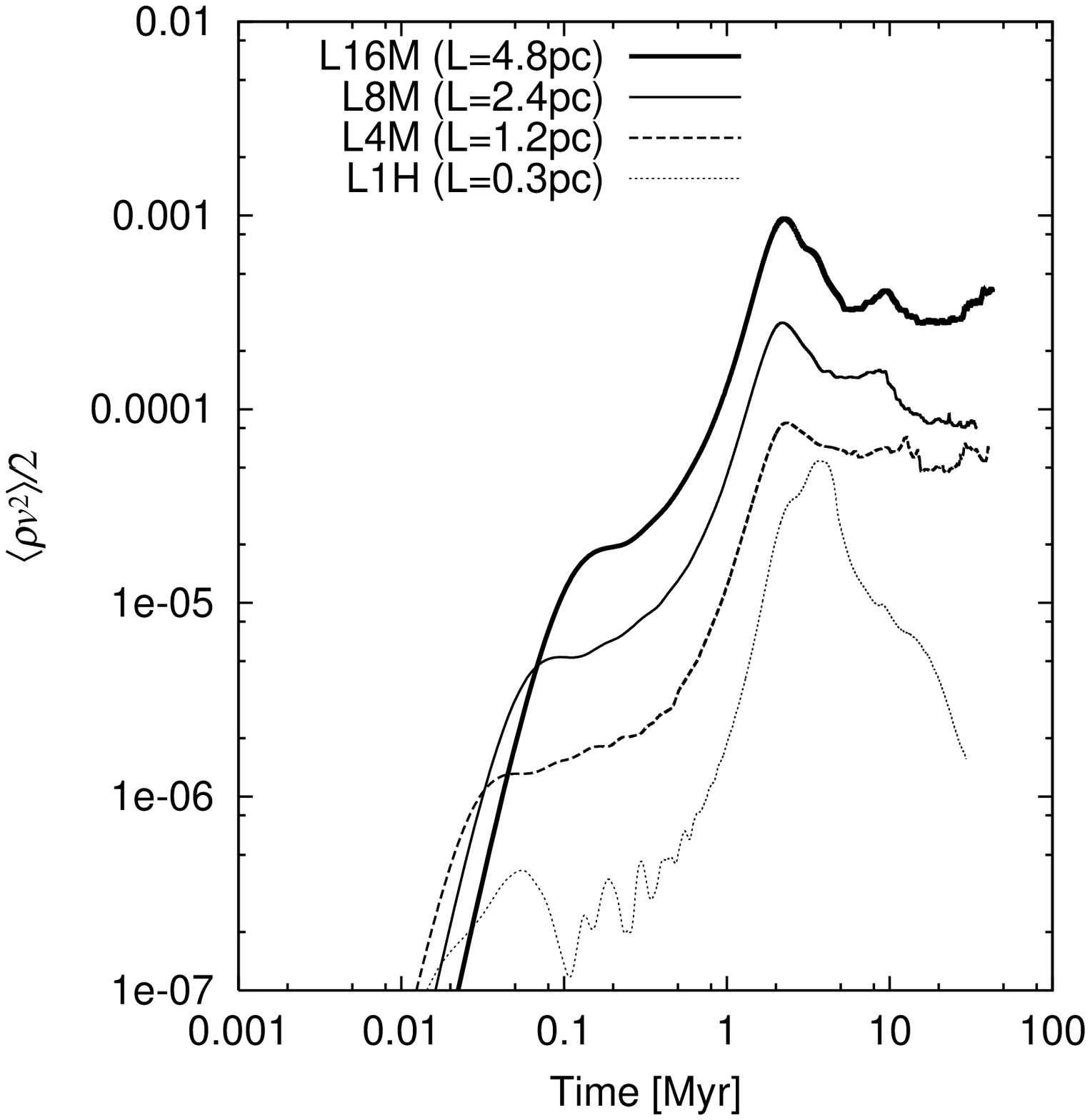}
\caption{Box size dependence: 
L=0.3 pc (L1H), 
1.2 pc (L4M), 
2.4 pc (L8M), 
4.8 pc (L16M). 
Pr=2/3 (see Table 1 for the initial parameters).}
 \label{size}
\end{figure}

Figure \ref{size} shows clearly that the saturation level of the turbulent 
 energy depends on the box size $L$:
 larger turbulent energy is obtained in the larger $L$ simulation.   
Model L1H ({\it dotted line}) having the box size of 0.3 pc
 shows decaying turbulence, 
 because the box size is so small that the long-wavelength mode can
 not develop.

\begin{figure}
\figurenum{7}
\epsscale{1.0}
\plotone{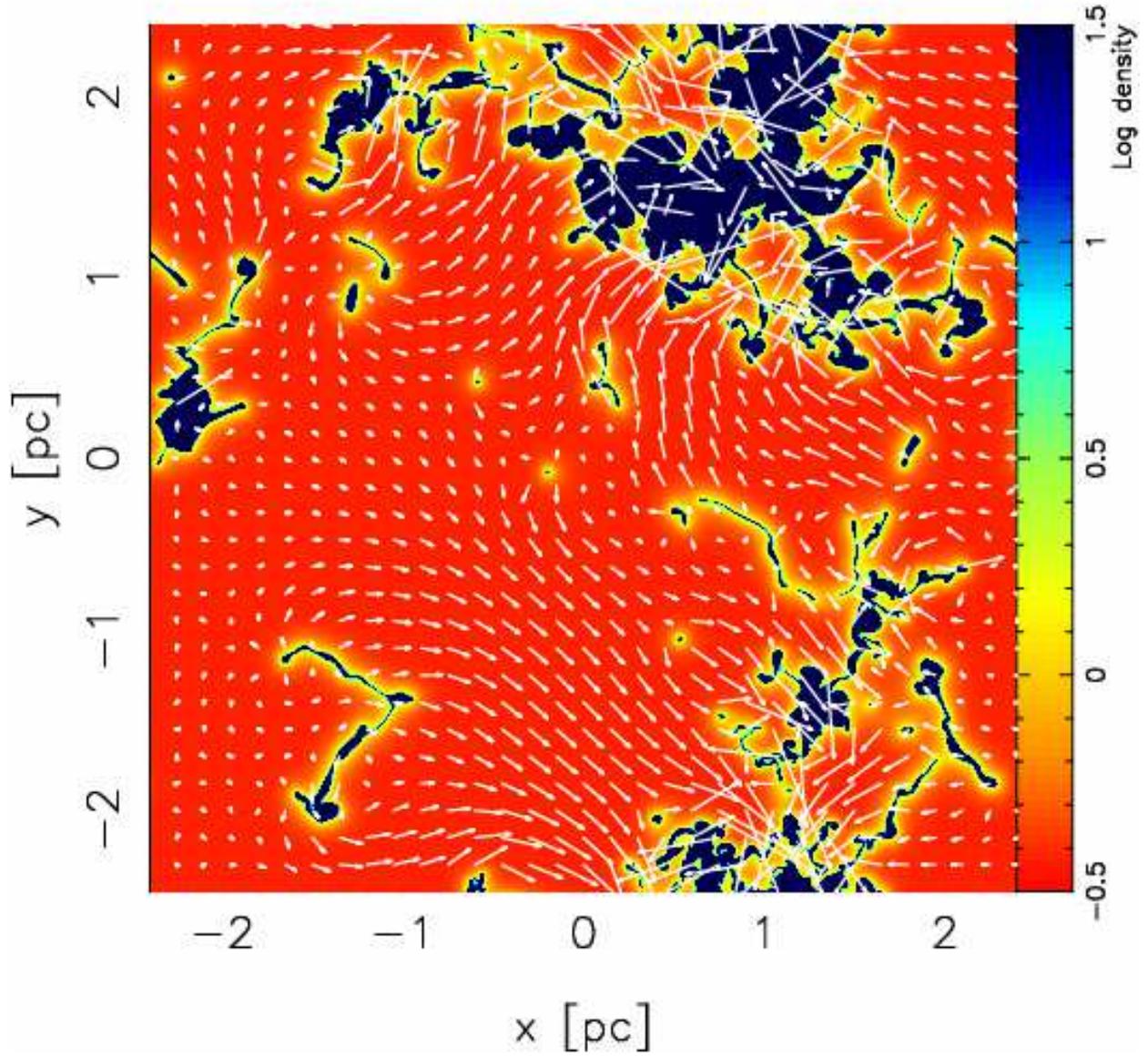}
\caption{Snapshot: Model L16M is shown (Pr=2/3, L=4.8pc).}
\label{L16}
\end{figure}

Figure \ref{L16} displays a snapshot of Model L16M at the late stage
 (see also thick line in Figure \ref{size}). 
The box size in this model has four times as large as in Model L4M
 (See Figure \ref{L4}).
The velocity dispersion of the model is 0.05 km/s which is still subsonic,
 indicating that the ratio 
 $L/\lambda_c=4.8/0.35=13.7$ is not sufficiently large to produce
 large velocity dispersion.
In order to satisfy the Field condition, 
 a large number of grids are basically
 required to cover long-wavelength, which means  
 high computational cost.
Two attempts to overcome this difficulty are
 described in the next subsection.

\subsection{Saturation Levels: Role of Long-Wavelength Mode}

In order to satisfy Field condition, the number of grids in
 one-dimension should not be less than $\sim L/\lambda_{\rm F}$.
Under this constraint, artificially large $\lambda_{\rm F}$
 can be used in order to reduce the grids.
Piontek \& Ostriker (2004) adopted similar approach with the large and
 temperature-independent conduction coefficient while we use the
 temperature-dependent coefficient, $K$.
The corresponding viscosity coefficient is adopted by fixing the Prandtl
 number to be 2/3. 
In most cases, the Field condition is moderately satisfied (see
 Table 2).

\begin{figure}
\figurenum{8}
\epsscale{1.0}
\plotone{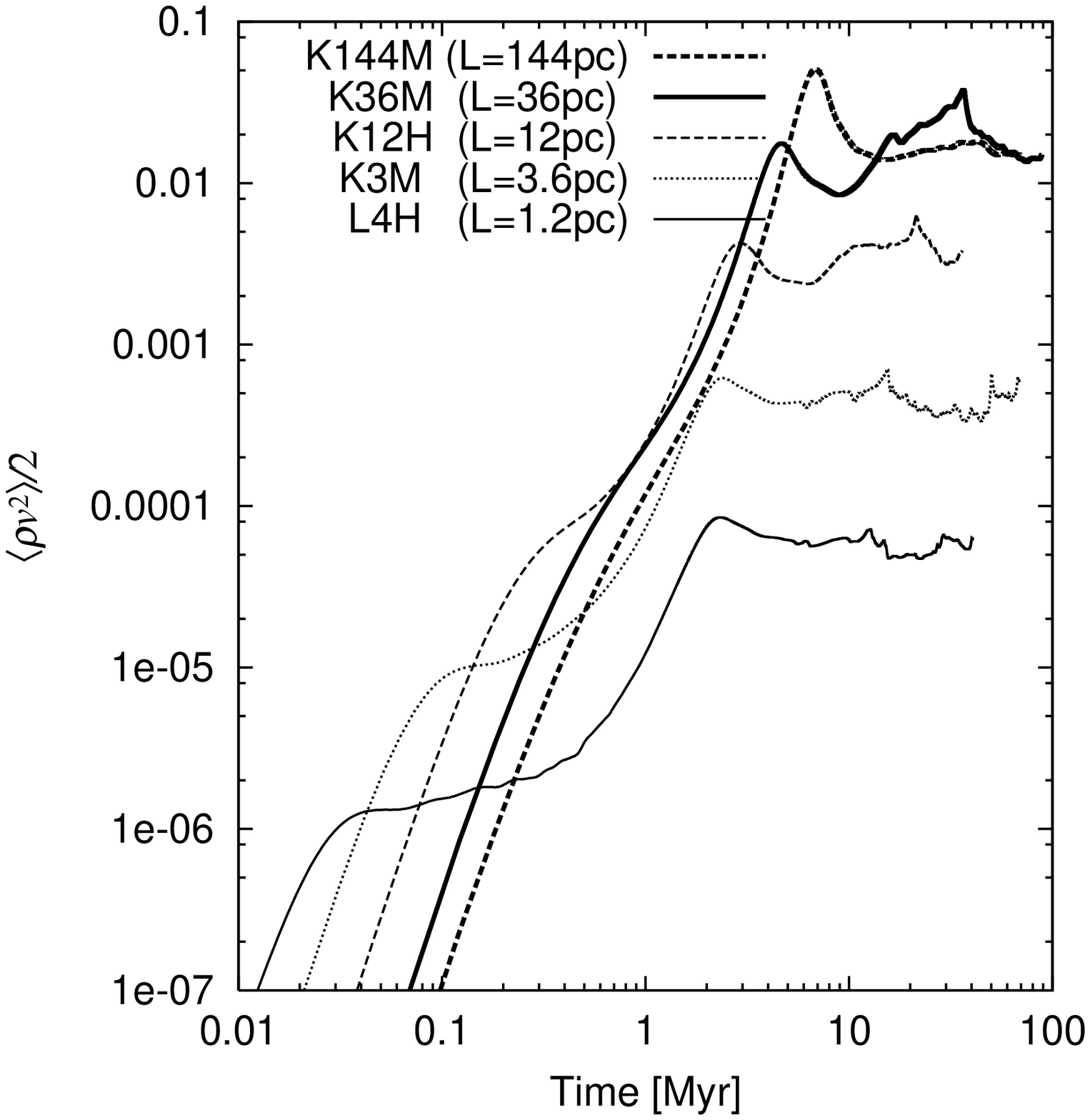}
 \caption{Box size dependence: 
Large conduction and viscosity coefficients are used in order to satisfy
 the Field condition. Pr=2/3. 
L=1.2 pc (L4H), 3.6 pc (K3M), 12 pc (K12H), 36 pc (K36M) and 144pc (K144M). 
} 
\label{sizeK}
\end{figure}
Figure \ref{sizeK} shows the results of larger box size simulations
 than that in Figure \ref{size},
 by using larger conduction and viscosity coefficients.
The larger domain simulation shows the larger saturation amplitude.
The Models K36M and K144M show the maximum saturation amplitude
 which is 0.3 km/s of the rms velocity.   
Large K run (K3M) and fiducial run (L16M) show almost the same
 amplitude.
This indicates that the amplitude does not depend only on K.

\begin{figure}
\figurenum{9}
\epsscale{0.3}
\plotone{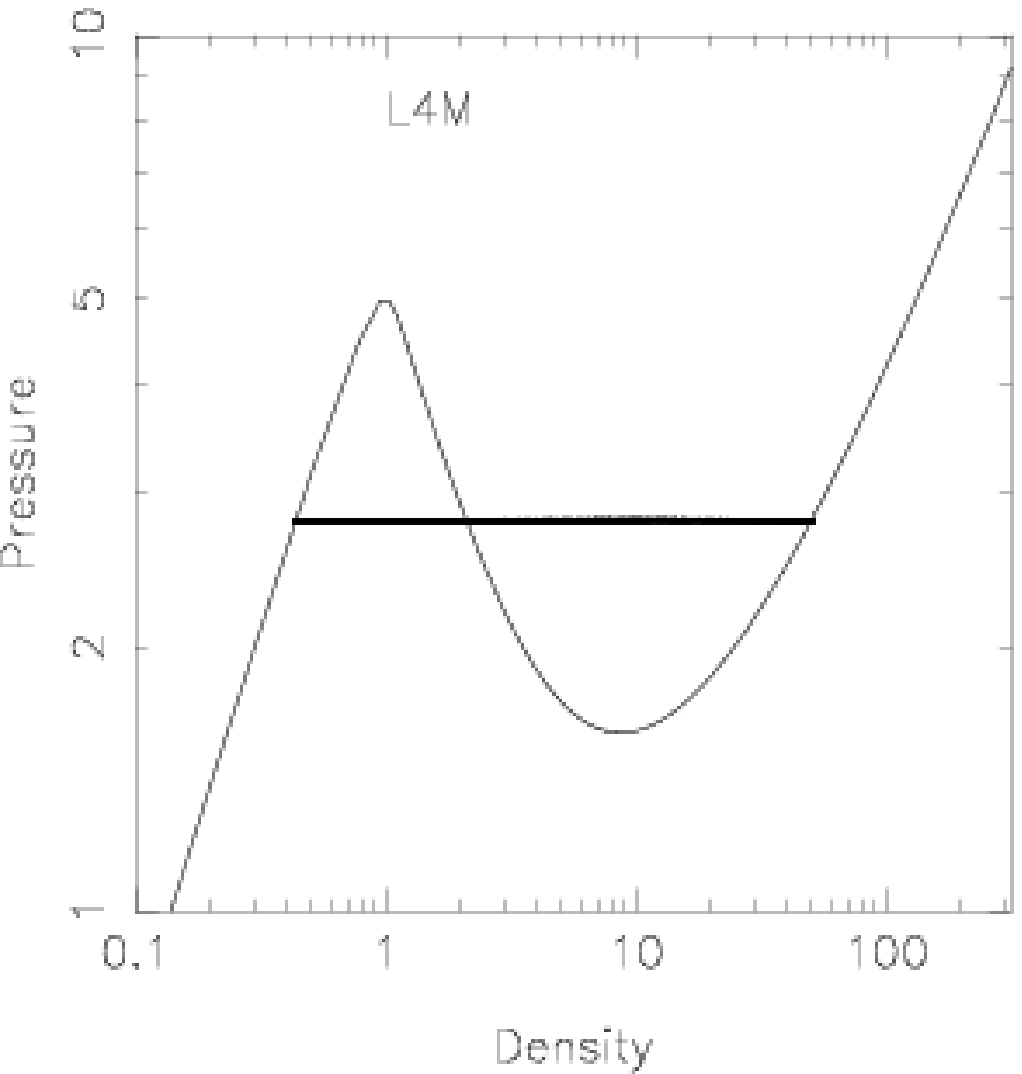}
\plotone{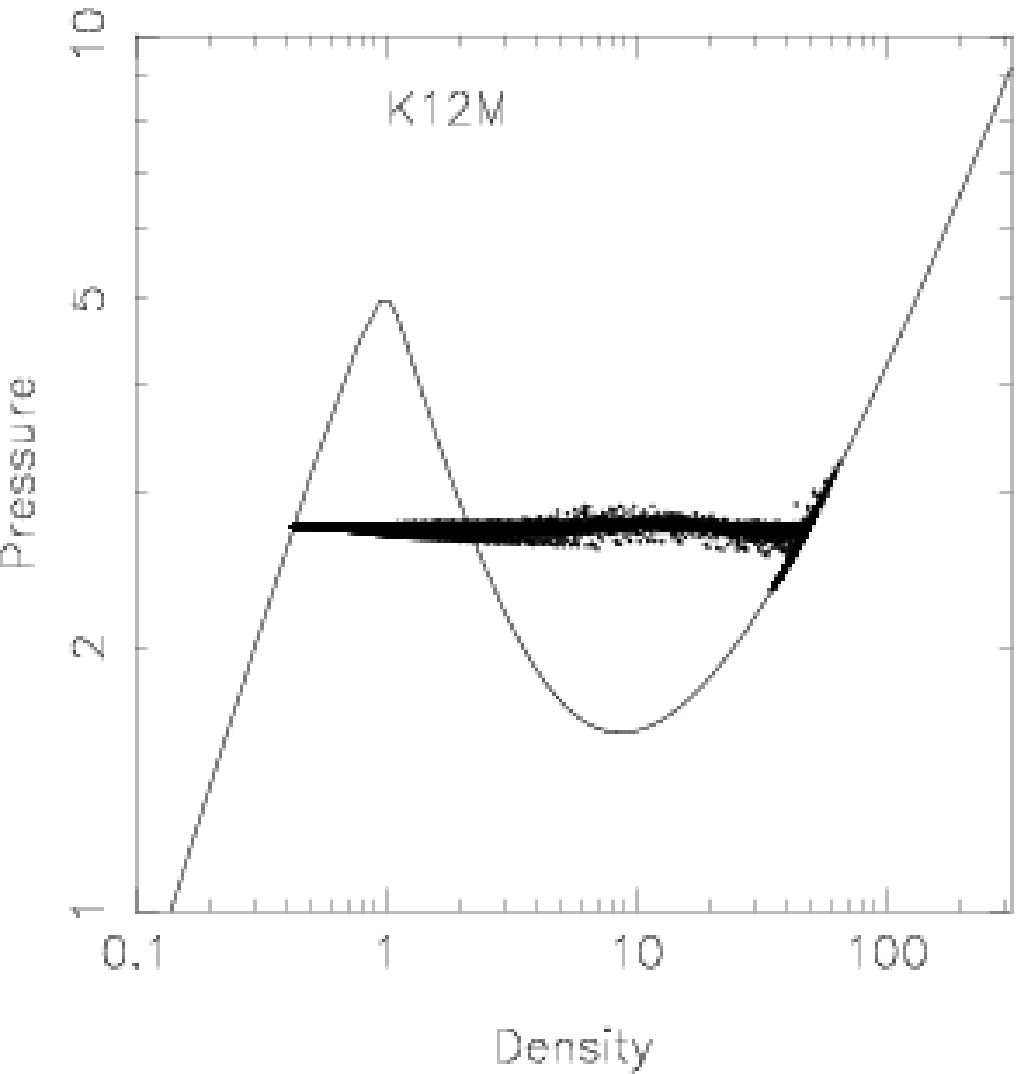}
\plotone{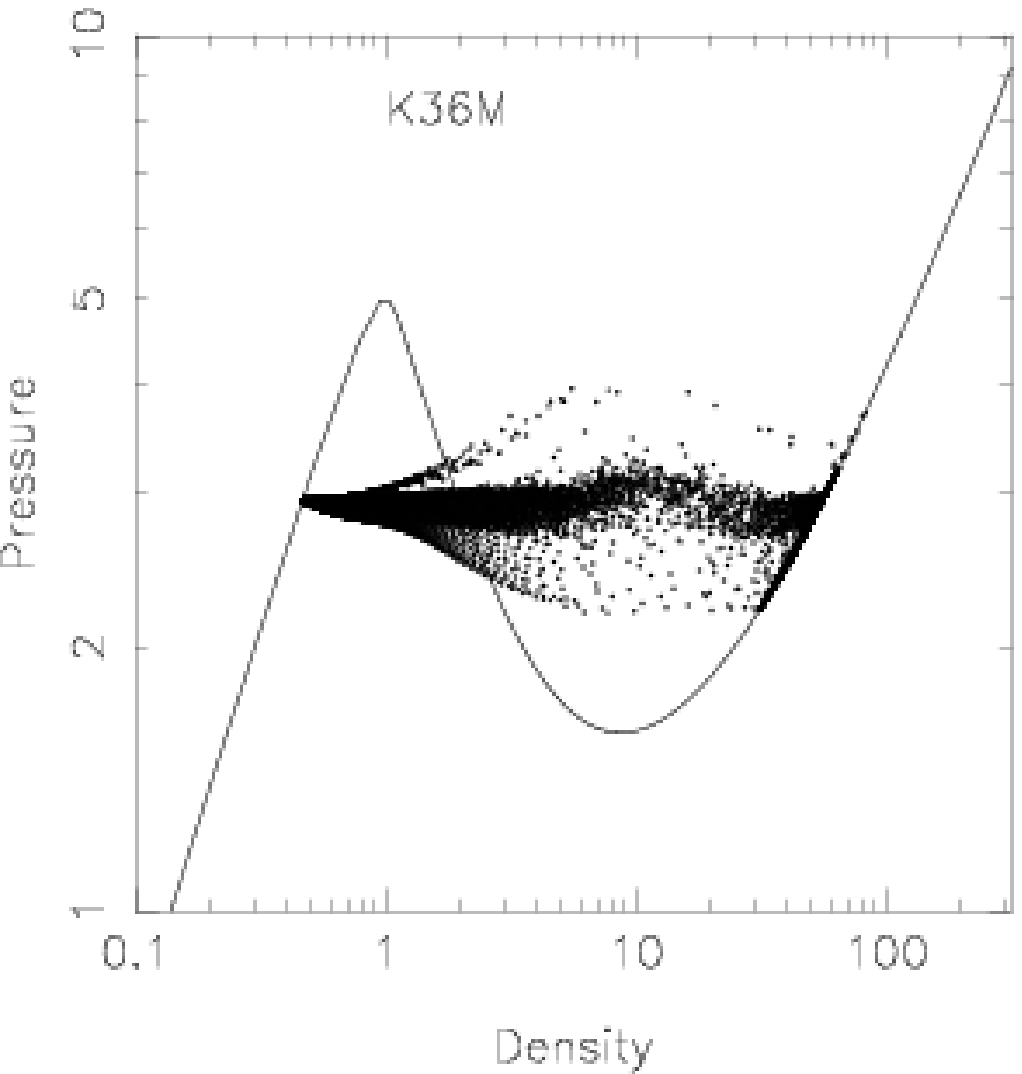}
\caption{
Pressure-density diagrams:
Pr=2/3, L=1.2 pc (L4M, {\it Left}), 
12 pc (K12M,{\it Middle}), 36 pc (K36M, {\it Right}). 
Typical late stage in each Model is plotted. 
Solid curves denote the
 thermal equilibrium state and the scatter dots are the fluid elements
 of all grids.}
\label{scatter} 
\end{figure}

We also present the pressure distribution in the three models (L4M, K12M,
 K36M) as a function of density (Figure \ref{scatter}). 
Almost uniform pressure distribution can be seen in the small box size
 simulation; 
 large scatter of pressure is seen in the large box run (e.g., K36M)
 indicating 
 that long-wavelength modes can generate large pressure gradients which
 consequently produce large turbulent velocities.

\begin{figure}
\figurenum{10}
\epsscale{1.0}
\plotone{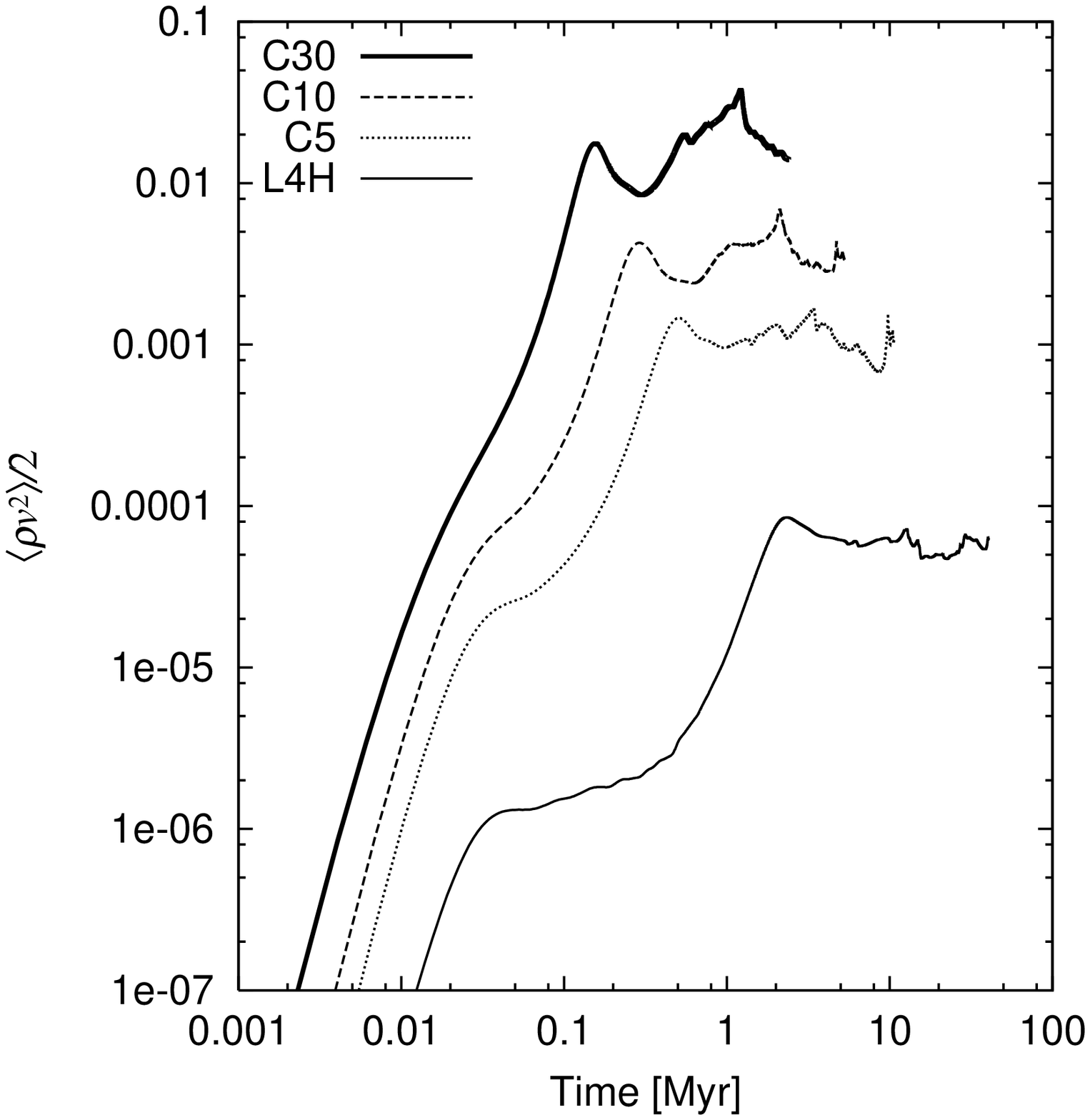}
 \caption{Saturation levels: Pr=2/3, L=1.2 pc. 
Models with different cooling rates are displayed (Models L4H, C5,
 C10, C30). }
\label{C5}
\end{figure}

Another attempt to simulate the long-wavelength mode is to reduce
 the cooling length $\lambda_c$ by using the large cooling rate $\Lambda$ 
 (Model C5, C10, C30, see Figure \ref{C5}).
In order to keep the same two-phase structure as that of fiducial model, 
 we take the same factor for $\Gamma$ and $K$, in the following form:
\begin{equation}
\frac{dE}{dt}= f \left[n\Gamma-n^2\Lambda-K\nabla^2 T\right]
\end{equation}
where $f$ is a constant number.
The results of these models show that the saturation
 amplitude depends on $\lambda_c^{-1}\propto f^{-1}$. 
As a conclusion, the larger amplitude are 
 obtained in both models (larger $L$ or smaller $\lambda_c$).   

\paragraph{Prandtl Number Dependence}

Section 4.3 has shown that the saturation is caused by the balance
 between TI and viscous dissipation.
This indicates that the larger amplitude will be obtained in
 simulations with the
 smaller viscosity and the larger thermal conduction coefficients.
Therefore, the ratio of viscosity and conduction may be a key
 parameter to determine the saturation level. 
To show this dependence we test the models with different Prandtl
 number.

We have measured the saturation velocity as a function of Prandtl number
 using a simple form, 
 $v\propto {\rm Pr}^{-m}\label{pr}$ and 
 obtained the exponent $m$=0.15 --- 0.5.
The uncertainty in $m$ is mainly due to small range of Pr, 
 $0\le {\rm Pr} \le 1$.

\begin{figure}
\figurenum{11}
\epsscale{1.0}
\plotone{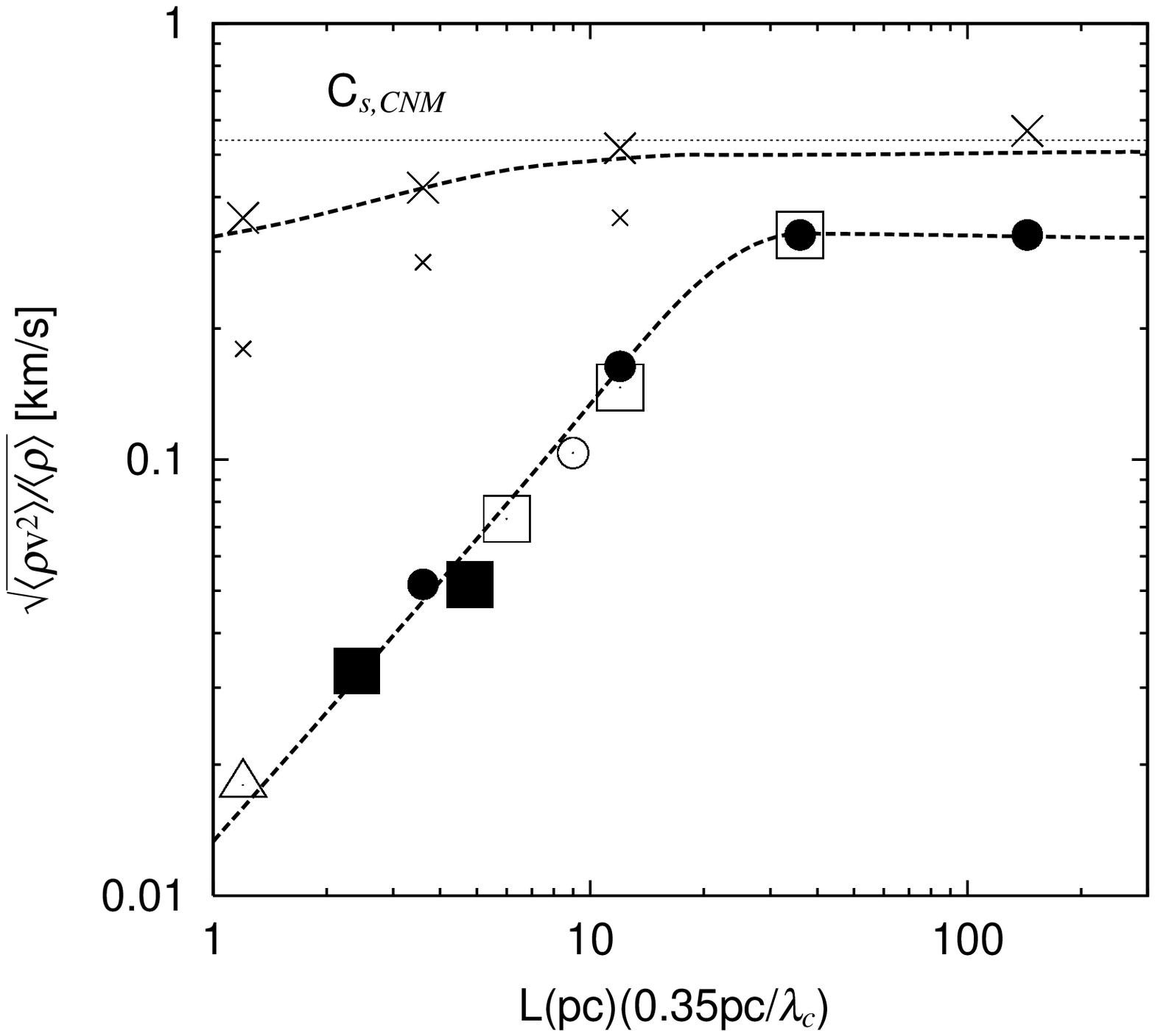}
 \caption{Saturation velocity vs length scale.
Open triangle: fiducial model (L4S). 
Filled boxes: Model L8, L16.
Filled circles: Model K9, K12, K36, K144.
Open circle: K9M3D (three-dimensional simulation).
Open boxes: Model C5, C10, C30.
Small crosses: Model K1P20, K3P20, K12P20 (left to right).
Large crosses: Model K1P100, K3P100, K12P100, K144P100 (left to right).}
\label{Scaling}
\end{figure}

The saturation amplitude as a function of the
 ratio $L/\lambda_c$ is plotted in Figure \ref{Scaling}. 
All the symbols except cross denote the results with Pr=2/3 showing
 a linear relation for $L/\lambda_c<100$ while they show saturation when
 $L/\lambda_c>100$.
In other words, at least $100\lambda_c$ in length is required to
 obtain the efficient development of TI.
Open circle denotes the result of the three-dimensional simulation (K9M3D).
We could not find any significant difference between two- and
 three-dimensional models 
% with respective to 
 in terms of
 the saturation amplitude.
The symbol cross denotes the result with smaller Prandtl number 
 which shows the greater amplitudes than that of Pr=2/3 models.

\section{Discussion}

\subsection{Comparison with Previous Works}

It is difficult to compare the saturation levels obtained here with 
 previous works 
 because almost all the works did not include physical
 viscosity, and the amount of the numerical
 viscosity is unknown. 
In fact, 
 Piontek \& Ostriker (2004) did not include physical viscosity 
 in their simulations that have
 shown similar saturated turbulence.
% with sufficient number of grids to resolve Field length. 
The model parameters they used are $L$ = 100 pc and $\lambda_c$=1.7 pc.
Simply assuming the Prandtl number to be 2/3, the velocity amplitude is
 expected to be 0.27 km/s from our results (Figure \ref{Scaling}) 
%(i.e., eqn. [\ref{eq}]).
while they measured 0.15 km/s for CNM.
There may be two possible reasons for this difference. 
One is the numerical Prandtl number of their code that was based on
 ZEUS-2D may be greater than 2/3.
The other is inefficient development of TI.
The Field length they used, $\lambda_{\rm F}=$3.125 pc was greater
 than $\lambda_{c}=$1.7 pc. 
This means that unstable modes whose wavelength $\sim \lambda_c$ 
 were suppressed owing to the thermal conduction.

Our results show that the smaller Pr simulations can achieve the
 larger saturation level.  
If the numerical method has spatially second-order accuracy, the numerical
 viscosity is supposed to scale as
 $\nu_{\rm numerical}\propto (\Delta x)^2$, and the numerical
 Prandtl number has the same dependence if the Field condition is
 satisfied. 
In this case, 
 the saturated velocity depends on $(\Delta x)^{-0.4}$ 
 where the exponent $m=0.2$ is assumed.
Thus, the increasing of the resolution by a factor of two 
 would give only 30 \% 
 increase in the velocity amplitude. 
Therefore, one tend to miss this small amplification in a
 numerical convergence test.

\subsection{Generation Mechanism of Turbulent Motions}

Physical process on the self-sustained turbulence
 demonstrated here may have fundamental importance as 
 a source of turbulence in ISM.
Each of CNM and WNM is thermally stable, having the similar
 character of decaying isothermal turbulence.
Here we emphasize an essential importance of  
 co-existence of CNM and WNM: 
 the contrast between one- to two-phase turbulence is more remarkable
 in the two-phase interface.  
Under an assumption of (one dimensional) plane-parallel geometry, 
 there is an unique equilibrium
 solution of CNM and WNM connected by the transition layer where the
 thermal conduction is important.
This equilibrium solution can be found by solving the energy equation
 $\nabla\cdot K\nabla T+n\Gamma-n^2\Lambda=0$ with the constant
 pressure that is the eigenvalue of the problem
 (Zel'dovich \& Pikelner 1969; Penston \& Brown 1970). 
Thus, the solution can be characterized by its pressure that is
 called ``saturation pressure.''
When the pressure of the system is not the same as the saturation
 pressure, there is no equilibrium solution for the static two-phase
 medium. 
If $P<P_{\rm sat}$, the gas flows across the interface from CNM to
 WNM, and vice versa.
The order of magnitude of this steady mass flow velocity is estimated by,
\begin{equation}
v\sim \frac{\lambda_{\rm F}}{\tau_{\rm cool}} 
=\frac{\gamma-1}{k_{\rm B}}\sqrt{\frac{K\Lambda}{T}}
\approx
0.0018T_{2}^{0.15} ~ {\rm km/s}.
\end{equation}
This velocity is lower than any saturation velocities obtained in this
 paper.
In other words, 
 approximate description based on the assumption of steady flow with
 infinite spatial extent is not sufficient to
 investigate self-sustained turbulence.
This indicates that 
 there might be some processes to amplify this steady-type flow.
One interesting mechanism for generating those translational motions is 
 the instability of a transition layer between WNM and CNM.
An analogous instability is known in a laminar
 flame propagation  
 (the so-called Darrieus-Landau instability).
The detailed analysis on this instability of transition layer is
shown in Inoue, Inutsuka \& Koyama (2006).

\section{Summary}

In this work we investigate the basic hydrodynamics 
 in interstellar two-phase medium by using two- and
 three-dimensional hydrodynamical simulations incorporating radiative 
 cooling/heating, thermal conduction, and physical viscosity but
 without any external forcing.  
Starting with small amplitudes of unstable modes of thermal 
 instability (TI),
 we calculate the nonlinear dynamics of the two-phase medium. 
Our findings are the following:

1. {\it Saturation Property}

In the fiducial model whose domain length is 1.2 pc,
 the kinetic energy produced by TI does not decay in at least 
 25 Myr which corresponds to 125 sound crossing time of WNM.
On the other hand, 
 the viscous dissipation timescale that is directly measured by
 the adopted viscosity term is about 6 Myr, 3 times greater
 than the sound crossing time of CNM. 
We should, therefore, conclude that the kinetic energy must be
 continuously supplied by some processes.
These kinetic motions may be produced by pressure gradients 
 which are caused by local inhomogeneous (density and temperature
 variation) cooling. 
As a conclusion, interstellar two-phase medium itself continuously produce
 turbulent motions. 
We call this dynamical equilibrium state ``saturation.''

2. {\it Saturation Amplitude}

The saturation is observed in most of the models
 whose domain lengths are greater than 0.3 pc.
The saturation amplitude is characterized by the ratio
 $L/\lambda_c$, where $L$ is the domain length and $\lambda_c$ is the
 cooling length defined by the product of the sound speed and 
 the cooling time.  
Decaying turbulence is found in the model with $L/\lambda_c<1$; 
Self-sustained turbulence is found in the model with $L/\lambda_c>1$.
The saturated velocity also show saturation when the ratio
 $L/\lambda_c$ is greater than 100. 
The maximum of the saturation velocity is 
 about a half of the sound speed of CNM
 if the Prandtl number is assumed to be 2/3. 

3. {\it Numerical Resolution for TI}

In order to obtain maximum saturation level of TI-induced turbulence, 
 the computational domain length needs to be greater than $100\lambda_c$. 
On the other hands, the Field length, the minimum characteristic
 lengths of TI, should be resolved to avoid resolution-dependent results.
Then, simulations should cover the length ranging from 
 $\lambda_{\rm F}$ to $100\lambda_c$. 
If the realistic parameters are used, the ratio 
 $\lambda_c/\lambda_{\rm F}$ is 117 and therefore 
 the total grids per one-dimension needs $\sim 10^4$
 which means prohibitively expensive calculation at
 present, in particular, for multi-dimensions.
Reducing the ratio $\lambda_c/\lambda_{\rm F}$ by using artificial
 parameters demonstrated in this paper is available.
However, one should meet the important criterion that 
 $\lambda_{\rm F}$ is less than $\lambda_c$
 in order not to suppress the most unstable mode of TI 
 by thermal conduction.

4. {\it Prandtl Number Dependence}

The saturation amplitude of turbulent velocity 
 does not change when we increase the
 viscosity and thermal conduction coefficients simultaneously 
 in order to keep the Prandtl number.
This shows that the thermal conduction plays an important role in
 maintaining turbulent motions against viscous dissipation.
The amplitude also increases when we decrease the Prandtl number:   
$v\propto$Pr$^{-m}$ with $m=$0.15 --- 0.5.

\acknowledgements

Numerical computations were carried out on VPP5000 (project ID:
 ysi04a, rhk26b and whk13b) at 
 the Astronomical Data Analysis Center of the National Astronomical
 Observatory, Japan (NAOJ). 
HK is supported by the 21st Century COE Program of Origin and
 Evolution of Planetary Systems in the Ministry of Education, Culture,
Sports, Science and Technology (MEXT) of Japan.
SI is supported by the Grant-in-Aid  (No.15740118, 16077202, 18540238)
from MEXT of Japan. 
%We thank the Astronomical Data Analysis Center of NAOJ for financial
% support for publication.

\singlespace

\end{document}